\documentclass[11pt]{article}

\usepackage[final]{acl}

\usepackage{amsmath}
\usepackage{enumitem}
\usepackage{amssymb}
\usepackage{booktabs}
\usepackage{caption}
\usepackage{subcaption}
\usepackage{multirow}
\usepackage{xcolor}
\usepackage{makecell}
\usepackage{graphicx}
\usepackage{tabularx}
\usepackage{booktabs}

\usepackage{times}
\usepackage{latexsym}

\usepackage[T1]{fontenc}

\usepackage[utf8]{inputenc}

\usepackage{microtype}

\usepackage{inconsolata}

\usepackage{graphicx}
\usepackage{xspace}
\usepackage[framemethod=tikz]{mdframed}
\newcounter{finding}
\newcommand{\finding}[1]{\refstepcounter{finding}
  \vspace{1.5mm}
 \begin{mdframed}[linecolor=gray,roundcorner=12pt,backgroundcolor=gray!15,linewidth=3pt,innerleftmargin=2pt, leftmargin=0cm,rightmargin=0cm,topline=false,bottomline=false,rightline = false]

  \textbf{Answer to RQ\arabic{finding}:} #1
 \end{mdframed}
 \vspace{1.5mm}
}

\title{EET: Experience-Driven Early Termination for Cost-Efficient \\Software Engineering Agents}

\author{Yaoqi Guo\textsuperscript{1}, Ying Xiao\textsuperscript{2}, Jie M.  Zhang\textsuperscript{2}, Mark Harman\textsuperscript{3}, Yiling Lou\textsuperscript{4}, \\\textbf{Yang Liu}\textsuperscript{1}, \textbf{Zhenpeng Chen}\textsuperscript{5}\thanks{Corresponding author: Zhenpeng Chen.}\\
        \textsuperscript{1}Nanyang Technological University    \textsuperscript{2}King's College London 
    \textsuperscript{3}University College London\\
  \textsuperscript{4}University of Illinois Urbana-Champaign
        \textsuperscript{5}Tsinghua University \\
        yaoqi001@e.ntu.edu.sg, \{ying.1.xiao, jie.zhang\}@kcl.ac.uk, mark.harman@ucl.ac.uk,\\ yilingl@illinois.edu, yangliu@ntu.edu.sg, zpchen@tsinghua.edu.cn
        }

\begin{document}
\maketitle

\newcommand{\gear}{EET\xspace}

\begin{abstract}
Software engineering (SE) agents powered by large language models are increasingly adopted in practice, yet they often incur substantial monetary cost. We introduce \gear, an experience-driven early termination approach that reduces the cost of SE agents while preserving task performance. \gear extracts structured experience from prior issue-resolution executions and leverages it to guide early termination during patch generation and selection, reducing unproductive iterations. We evaluate \gear on the SWE-bench Verified benchmark across three representative SE agents. \gear consistently reduces total cost by 19\%–55\% (32\% on average), with negligible loss in resolution rate (at most 0.2\%). These efficiency gains are achieved, on average, by identifying early-termination opportunities for 11\% of issues and reducing API calls, input tokens, and output tokens by 21\%, 30\%, and 25\%, respectively. We release the code, prompts, and data at \url{https://github.com/IanWalls/EET}.
\end{abstract}

\section{Introduction}

Large language models (LLMs) are reshaping modern software development, with software engineering (SE) agents emerging as one of their most impactful applications~\cite{seagentsurvey,rahardja2025can}. Given an issue describing a bug or a feature request along with the corresponding code repository, an SE agent aims to navigate the codebase, localize relevant code, and generate a patch (i.e., modify the code) to resolve the issue, with correctness validated by the associated tests~\cite{xia2024agentless}.

To achieve this, a variety of SE agents have been developed. For example, Agentless~\cite{xia2024agentless} follows a fixed, expert-designed workflow without autonomous planning. In contrast, Mini-SWE-Agent~\cite{yang2024sweagent} and Trae Agent~\cite{gao2025trae} are autonomous, capable of interacting with the environment, using tools, and planning multi-step actions. 

Despite the remarkable capabilities of existing SE agents, cost efficiency remains a key concern for practical adoption. In a recent Stack Overflow survey~\cite{sosurvey}, 53\% of software engineers reported that the cost of using agents is a barrier for them. This monetary cost is also associated with environmental impact due to energy consumption~\cite{ren2024reconciling} and other resource usage~\cite{jegham2025hungryaibenchmarkingenergy}. Reducing cost inefficiency is therefore important for both economic and environmental reasons.

This cost inefficiency arises largely because SE agents typically resolve issues through multi-round iterations~\cite{gao2026more}, such as repeated tool invocations. In each iteration, the entire conversation history, including prompts, tool calls, and intermediate outputs, is fed back as context, causing computational costs to grow super-linearly, a phenomenon known as ``token snowball'' in SE agents~\cite{abs250909853}. Moreover, when agents are stuck on difficult or unsolvable issues, they often continue consuming iterations that provide little additional benefit to successful issue resolution, further amplifying unnecessary cost~\cite{abs250909853}.

To address this challenge, we propose \gear, an experience-driven early termination approach for cost reduction of SE agents while preserving task performance. 
The key insight behind \gear is that, much like an experienced developer, an agent informed by relevant prior experience can navigate directly to a solution, bypassing the exhaustive and costly iterations typical of `trial-and-error' approaches.

\gear realizes this through two complementary mechanisms. First, it captures structured experience from historical issue-resolution activities, encoding information about the issue, the agent’s execution trajectory, and the outcomes of prior attempts. Second, when tackling a new issue, \gear retrieves relevant experience to guide the agent’s behavior, enabling early termination of redundant iterations during both patch generation and selection. By leveraging past experience, \gear reduces unnecessary computational cost while maintaining or even improving the likelihood of successfully resolving the issue. Importantly, \gear is a highly general optimization approach that can be integrated seamlessly into diverse agents without requiring fundamental redesigns.

We evaluate \gear on the widely adopted SWE-bench Verified benchmark~\cite{openai2025swebenchverified} using three representative SE agents, namely Agentless~\cite{xia2024agentless}, Mini-SWE-Agent~\cite{yang2024sweagent}, and Trae Agent~\cite{gao2025trae}, across different LLM backends. The results show that \gear consistently reduces total cost by 19.3\%–55.1\% (31.8\% on average), with negligible loss in resolution rate (at most 0.2\%). These efficiency gains are achieved by identifying early-termination opportunities for 8.6\%–14.0\% of issues (11.3\% on average), as well as by reducing API calls, input tokens, and output tokens by 20.8\%, 29.9\%, and 25.1\% on average, respectively.
\section{Related Work}
\noindent \textbf{SE Agents.}
SWE-bench~\cite{jimenez2024swebench} and its human-validated variant SWE-bench Verified~\cite{openai2025swebenchverified} have become standard benchmarks for evaluating automated resolution of real-world SE issues. Motivated by these benchmarks, recent work has focused on agent-based approaches that equip LLMs with tool use and iterative execution to autonomously explore codebases and generate patches~\cite{seagentsurvey}. Representative examples include Mini-SWE-Agent~\cite{yang2024sweagent}, which relies on shell-based interaction for iterative code navigation and editing, and Trae Agent~\cite{gao2025trae}, which combines parallel patch generation with a selector agent that leverages static analysis and execution traces.
While these agents achieve strong performance, previous work~\cite{abs250909853} has shown that such performance is accompanied by substantial token consumption.

\noindent \textbf{Agent Efficiency Optimization.} Previous work on improving agent efficiency can be grouped into three categories. First, prompt-level approaches reduce input overhead through compression~\cite{jiang2023llmlingua} or Retrieval-Augmented Generation (RAG)~\cite{ shinn2023reflexion}. Second, agent re-architectures lower cost via strategies such as plan caching~\cite{zhang2025cost}, codified prompting~\cite{yang2025codeagents}, budget-aware structuring~\cite{yang2026bamas}, or multi-agent delegation~\cite{gandhi2024budgetmlagent}. Third, reasoning constraints control runtime resource use, e.g., per-turn token budgeting~\cite{han2024token}, static safeguards~\cite{jimenez2024swebench}, or dynamic turn-control~\cite{gao2026more}. While effective at reducing cost, these approaches typically degrade task performance, highlighting the challenge of achieving both efficiency and performance.

\noindent \textbf{Agent Memory.}
Recent agents incorporate memory mechanisms to improve task performance, such as shared message pools in MetaGPT~\cite{hong2024metagpt} and hierarchical memory designs in Generative Agents~\cite{park2023generative} and MemoryBank~\cite{zhong2024memorybank}. More broadly, RAG~\cite{lewis2020retrieval} extends agent context via external memory. The experience used in \gear can be viewed as a form of agent memory. However, previous work primarily uses memory to improve task performance and often incurs additional computational cost due to frequent retrieval and context expansion. For example, MemoryBank~\cite{zhong2024memorybank} requires frequent memory retrieval operations, leading to increased cost. In contrast, \gear leverages structured experience specifically to reduce agent cost.

\begin{figure*}
    \centering
    \includegraphics[width=0.95\linewidth]{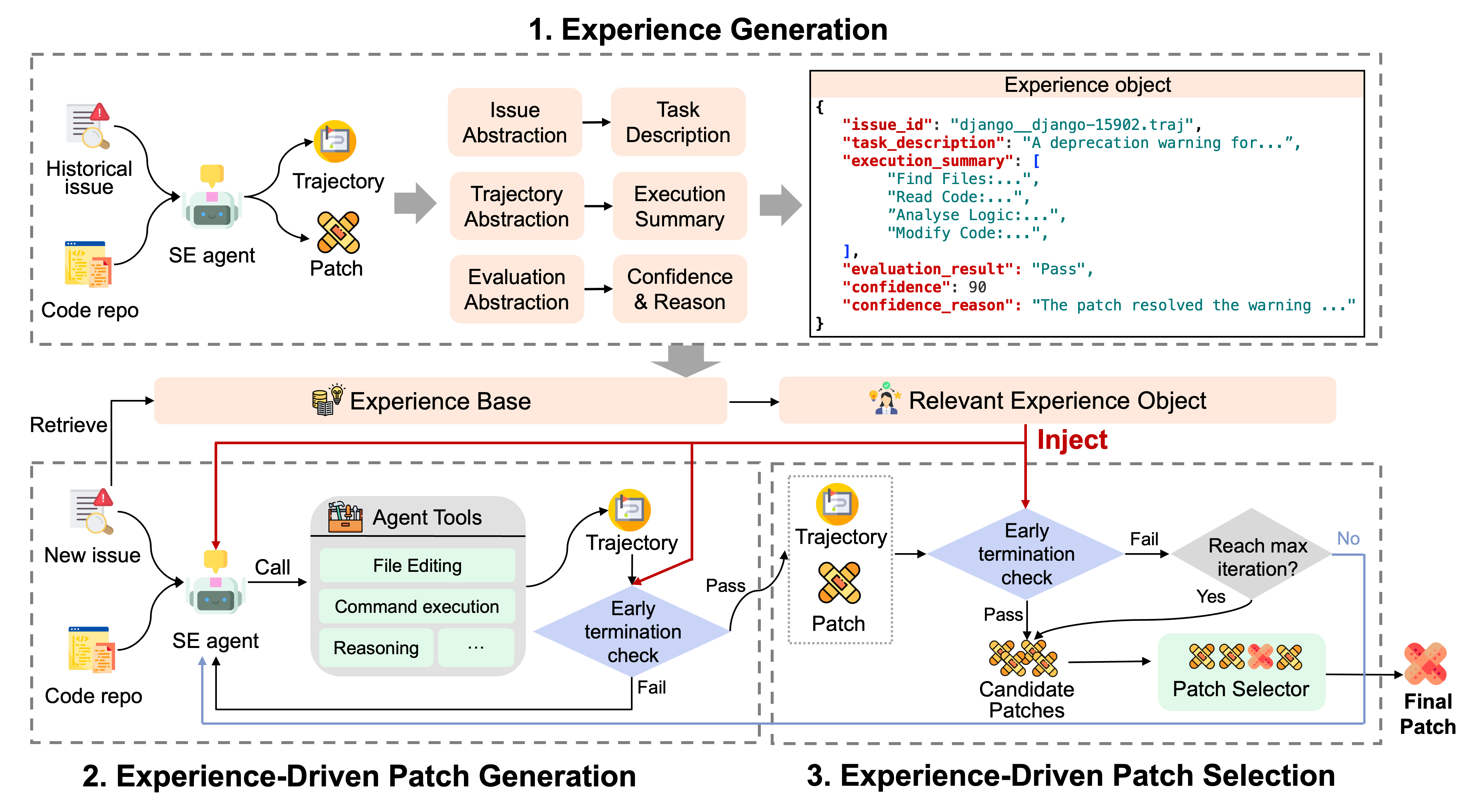}
    \caption{Overview of \gear.}
    \label{fig:mainMethod}
\end{figure*}

\section{Methodology}
Figure~\ref{fig:mainMethod} presents an overview of \gear, our experience-driven early termination approach for improving the cost efficiency of SE agents. 

Specifically, \gear consists of two components: (1) experience generation and (2) early termination of patch generation and selection. Experience generation records agent execution trajectories, i.e., sequences of tool calls and their outcomes, together with resolution results from historical issues, and summarizes them into an experience base. For a new issue, the SE agent retrieves relevant experience objects during patch generation to guide tool usage and enable early stopping of unproductive iterations. Once a patch and its execution trajectory are passed to the patch selection stage, \gear evaluates them against retrieved experience to both guide efficient selection and determine whether further patch generation is necessary.

\subsection{Experience Generation}
\label{chapter:experience_generation}
The experience generation component aims to distill an agent’s historical issue resolution experience into an experience base that can support the resolution of new issues. During issue resolution, agents produce complex execution trajectories consisting of tool calls and their execution outcomes, which serve as the primary source of experience. However, storing raw trajectories introduces substantial noise and token overhead, while overly aggressive simplification may discard actionable signals. We therefore design a balanced experience representation that preserves decision-relevant information while remaining compact and efficient.

\noindent \textbf{Experience Representation.} 
Figure~\ref{fig:mainMethod} illustrates an example schema of an experience object (a complete example is provided in Appendix \ref{appendix:experience_example}). Each experience object captures an agent’s experience from resolving a single issue and is tagged with the corresponding issue ID. The remaining fields provide a high-level abstraction of the issue resolution task, the resolution process, and the resolution outcome, enabling the experience to generalize across related issues. This abstraction is performed using the agent’s backend LLM.
 
In the following, we describe each field in detail.

\textbf{Issue abstraction:} Real-world issue descriptions are often lengthy and detailed~\cite{jimenez2024swebench}, containing background information, discussion, and auxiliary context that is not directly relevant to resolution. For efficiency, we store a concise abstraction of each issue in the \texttt{task\_description} field, capturing its core intent and technical requirements. This abstraction serves as a stable semantic anchor for matching and retrieving relevant experience for future issues.
    
\textbf{Trajectory abstraction:} Rather than recording every tool invocation and execution detail, we compress each execution trajectory into a sequence of key resolution steps, storing in the \texttt{execution\_summary} field. The abstraction process filters out non-essential intermediate actions and retains only critical steps that influence the final outcome, significantly reducing token usage compared to storing full trajectories.

\textbf{Evaluation abstraction:} Issue resolution typically includes a patch evaluation phase, which validates whether a generated patch resolves the issue by executing tests. In \gear, we retain only high-quality experience, namely resolution trajectories that lead to a successfully validated patch. From an efficiency perspective, unsuccessful attempts primarily reflect exploratory trial-and-error and provide limited reusable value, while successful resolutions capture actionable signals that can directly inform early stopping and decision-making. Accordingly, the \texttt{evaluation\_result} field is set to \emph{pass} for all stored experiences.

To further distill high-level and reusable knowledge from these resolution experiences, we introduce two additional fields: \texttt{confidence}, which provides a quantitative estimate of the reliability of the resolution process, and \texttt{confidence\_reason}, which offers a structured retrospective assessment of the generated patch along predefined criteria, including completeness, quality, relevance, and supporting evidence. Together, these fields form lightweight quantitative and qualitative abstractions of experience. By compressing raw execution traces into structured, high-value knowledge, they can serve as an efficiency driver that enables agents to bypass redundant trial-and-error and terminate unproductive iterations earlier.

\noindent \textbf{Experience Retrieval.} 
Experiences collected from historical issue resolutions are stored in an experience base. To enable their effective reuse in future issue resolution, the experience base supports experience retrieval. Given a new issue, we retrieve relevant experience objects by applying a commonly used text similarity approach TF–IDF~\cite{salton1988term} to match the current issue description against the \texttt{task\_description}s stored in the experience base.

The retrieval strategy is designed to balance both task performance and efficiency. To preserve task performance, we mitigate context pollution, where irrelevant experiences waste tokens and distract the agent, by enforcing a similarity threshold $\tau_{\text{sim}}$. To ensure efficiency, we retrieve only the top-1 experience whose similarity to the current issue exceeds this threshold. This ensures that the agent is provided with a highly relevant, compact, and high-quality reference with minimal overhead.

\subsection{Early Termination}
Existing SE agents typically terminate patch generation and selection only when the backend LLM explicitly signals completion or when a predefined iteration budget is exhausted. Such coarse-grained control often leads to redundant computation, either by over-solving simple issues or by persisting on difficult ones with little chance of success.

\gear addresses this inefficiency through an early termination strategy. The key insight is that, when interpreted in the context of relevant historical experience, intermediate execution signals, such as code modifications, test feedback, and execution trajectories, often provide sufficient evidence to determine whether further iterations are necessary. Leveraging structured experience from past issue resolutions, \gear enables SE agents to make informed, dynamic stopping decisions during both patch generation and patch selection.

\subsubsection{Early Termination of Patch Generation}
During patch generation, SE agents iteratively produce a patch by interacting with the code repository and execution environment through tool invocations such as file editing and command execution. These interactions and their outcomes are recorded as an execution trajectory.

Rather than relying on explicit termination signals from the backend LLM or predefined iteration limits, \gear augments this process with experience-driven guidance and early termination. For a new issue, \gear retrieves relevant experience objects and conditions patch generation on both the issue description and the retrieved experience. More importantly, it introduces early termination checks at intermediate milestones during patch generation.

Specifically, \gear treats both code modification and test execution as milestones. After each milestone, the agent estimates a confidence score indicating whether the current modifications are sufficient to resolve the issue, informed by retrieved experience. If the confidence score exceeds a threshold $\tau^{\text{gen}}$, patch generation is terminated early; otherwise, the agent continues iterating. Patch generation is also terminated when the maximum iteration budget of the underlying SE agent is reached.

This dual-milestone design reflects the practical observation that confidence in a patch may arise either immediately after code modifications, based on structural and semantic alignment with past successful fixes, or after observing dynamic feedback from test executions. By enabling early termination at both stages, \gear avoids redundant iterations while preserving patch quality. 

\subsubsection{Early Termination of Patch Selection}
\label{subsec:methdology_patch_selection}
To account for the stochasticity of LLMs, SE agents commonly generate multiple candidate patches and select among them using reproduction and regression tests. In practice, the number of generated patches is often fixed by a maximum iteration limit, leading to inefficiencies when additional patches are unlikely to improve outcomes.

\gear introduces an early termination check to dynamically control the number of generated patches. After each patch is generated, \gear evaluates its potential using three inputs: the patch itself, the corresponding execution trajectory, and relevant experience retrieved from past issue resolutions. Based on these inputs, the backend LLM is prompted to produce a confidence score indicating the likelihood that the patch will lead to a successful resolution. To improve reliability and interpretability, the LLM is also required to provide a brief rationale for its confidence assessment.

Early termination decisions are made by comparing the confidence score against two thresholds, $\tau_{\text{upper}}^{\text{sel}}$ and $\tau_{\text{lower}}^{\text{sel}}$. If the score exceeds $\tau_{\text{upper}}^{\text{sel}}$, the patch is deemed sufficiently reliable, and further patch generation is terminated to eliminate redundancy. Conversely, if the score falls below $\tau_{\text{lower}}^{\text{sel}}$, the issue is considered unlikely to be resolved by the current agent configuration, and additional resource-intensive iterations are halted to avoid waste. In both cases, the accumulated patches are passed to the patch selector to produce the final output.
If the confidence score lies between the two thresholds and the maximum iteration limit has not been reached, the agent proceeds to generate additional patches. Otherwise, once the iteration limit is reached, all accumulated patches are forwarded to the patch selector.

Unlike the early termination check during patch generation, which avoids unnecessary cost within the generation of a single patch, experience-driven patch selection prevents wasteful generation of additional patches. Together, these two complementary early termination checks enable \gear to substantially improve the cost efficiency of SE agents without compromising task performance.

\section{Evaluation Setup}

\subsection{Research Questions (RQs)}

We aim to evaluate \gear by answering three RQs.

\noindent\textbf{RQ1 (Effectiveness of \gear):} What is the impact of \gear on task performance and cost?

\noindent\textbf{RQ2 (Comparison with baselines):} How does \gear compare with baseline methods?

\noindent\textbf{RQ3 (Ablation study):} How do experience injection and early termination individually affect \gear's effectiveness?

\subsection{Datasets} 
We evaluate \gear on SWE-bench Verified~\cite{openai2025swebenchverified}, a human-validated benchmark released by OpenAI for reliably assessing AI models’ ability to resolve SE issues. It has been the most widely adopted dataset for evaluating SE agents~\cite{swebench2025}. The dataset consists of 500 real-world tasks collected from Python repositories on GitHub. Each task is derived from a resolved GitHub issue and is accompanied by a set of unit tests used to validate candidate patches.

For each task, agents are provided with the original GitHub issue description as the problem statement and granted access to the corresponding code repository. Based on this information, agents are required to modify the repository files to fix the issue. The unit tests are withheld from the agents and are used solely for evaluation.

To generate experience for \gear, we additionally use SWE-bench Lite~\cite{jimenez2024swebench}, a commonly used auxiliary dataset that contains 300 tasks following the same format as SWE-bench Verified. To prevent data leakage, we remove any tasks that overlap with SWE-bench Verified. After deduplication, 207 tasks remain, which are used exclusively for experience generation.

\begin{table*}[t]
\small
  \centering
  \definecolor{goodblue}{HTML}{0077BB}
  \definecolor{badorange}{HTML}{EE7733}

  \begin{tabular}{lrrrrr}
    \toprule
    \textbf{Agent} & \textbf{\% Resolved} & \textbf{\# API Calls} & \textbf{\# Input Tokens} & \textbf{\# Output Tokens} & \textbf{Total Cost} \\
    \midrule
    \multicolumn{6}{c}{\textbf{GPT-5-mini}} \\
    \midrule
    Agentless & 33.2\% & 4,500 & 35,919,315 & 1,861,379 & \$13.77 \\
    \quad \textit{with \gear} & 41.0\% & 3,310 & 17,306,929 & 912,076 & \$6.18 \\
    \quad \textit{Change} & \textcolor{goodblue}{+7.8\%} & \textcolor{goodblue}{-26.4\%} & \textcolor{goodblue}{-51.8\%} & \textcolor{goodblue}{-51.0\%} & \textcolor{goodblue}{-55.1\%} \\
    \midrule
    Mini-SWE-Agent & 55.6\% & 7,250 & 87,818,230 & 3,775,009 & \$16.07 \\
    \quad \textit{with \gear} & 56.6\% & 6,679 & 75,766,697 & 3,635,735 & \$12.96 \\
    \quad \textit{Change} & \textcolor{goodblue}{+1.0\%} & \textcolor{goodblue}{-7.9\%} & \textcolor{goodblue}{-13.7\%} & \textcolor{goodblue}{-3.7\%} & \textcolor{goodblue}{-19.4\%} \\
    \midrule
    Trae Agent & 66.2\% & 58,607 & 1,802,021,581 & 23,187,391 & \$161.79 \\
    \quad \textit{with \gear} & 66.2\% & 41,086 & 1,254,073,114 & 16,686,359 & \$116.18 \\
    \quad \textit{Change} & \textcolor{gray}{0.0\%} & \textcolor{goodblue}{-29.9\%} & \textcolor{goodblue}{-30.4\%} & \textcolor{goodblue}{-28.0\%} & \textcolor{goodblue}{-28.2\%} \\
    \midrule
    \multicolumn{6}{c}{\textbf{DeepSeek-V3.2}} \\
    \midrule
    Agentless & 42.6\% & 4,500 & 36,996,894 & 1,898,606 & \$11.21 \\
    \quad \textit{with \gear} & 49.8\% & 3,351 & 25,183,691 & 1,234,093 & \$7.60 \\
    \quad \textit{Change} & \textcolor{goodblue}{+7.2\%} & \textcolor{goodblue}{-25.5\%} & \textcolor{goodblue}{-31.9\%} & \textcolor{goodblue}{-35.0\%} & \textcolor{goodblue}{-32.2\%} \\
    \midrule
    Mini-SWE-Agent & 66.4\% & 33,896 & 614,853,823 & 6,696,703 & \$52.20 \\
    \quad \textit{with \gear} & 67.0\% & 31,054 & 531,220,952 & 6,402,107 & \$42.15 \\
    \quad \textit{Change} & \textcolor{goodblue}{+0.6\%} & \textcolor{goodblue}{-8.4\%} & \textcolor{goodblue}{-13.6\%} & \textcolor{goodblue}{-4.4\%} & \textcolor{goodblue}{-19.3\%} \\
    \midrule
    Trae Agent & 70.2\% & 151,038 & 4,979,212,134 & 39,770,353 & \$170.01 \\
    \quad \textit{with \gear} & 70.0\% & 111,039 & 3,102,450,888 & 28,550,220 & \$107.56 \\
    \quad \textit{Change} & \textcolor{badorange}{-0.2\%} & \textcolor{goodblue}{-26.5\%} & \textcolor{goodblue}{-37.7\%} & \textcolor{goodblue}{-28.2\%} & \textcolor{goodblue}{-36.7\%} \\
    \midrule
    \midrule
    \textbf{Average Change} & \textbf{\textcolor{goodblue}{+2.7\%}} & \textbf{\textcolor{goodblue}{-20.8\%}} &  \textbf{\textcolor{goodblue}{-29.9\%}}  & \textbf{\textcolor{goodblue}{-25.1\%}}  & \textbf{\textcolor{goodblue}{-31.8\%}} \\
    \bottomrule
  \end{tabular}
  \caption{(RQ1) Comparison of task performance and cost of SE agents with and without \gear. Changes in \%~Resolved are reported as absolute differences, while changes in other metrics are computed as relative changes.}
  \label{tab:mainResults}
\end{table*}

\subsection{SE Agents}\label{agentdes}
We evaluate \gear on three representative, state-of-the-art open-source SE agents: Agentless~\cite{xia2024agentless}, Mini-SWE-Agent~\cite{yang2024sweagent}, and Trae Agent~\cite{gao2025trae}. They cover three different paradigms of SE agent design. Agentless follows a fixed, expert-designed workflow without autonomous planning or complex tool use, while Mini-SWE-Agent and Trae Agent are autonomous agents capable of tool use, environment interaction, and multi-step planning. Mini-SWE-Agent directly generates a single patch, whereas Trae Agent performs both patch generation and selection. 

Because Agentless follows a predefined, fixed workflow for patch generation and does not support autonomous tool invocation, \gear cannot be applied to its patch generation phase and is instead applied only to patch selection. In contrast, Mini-SWE-Agent does not include a patch selection phase; therefore, \gear is applied solely during patch generation. For Trae Agent, which supports both patch generation and selection, \gear is applied to both phases.

For each agent, we implement two variants using GPT-5-mini and DeepSeek-V3.2 as the LLM backends. These two models are widely adopted representatives of closed-source and open-source LLMs, respectively. All agents are implemented with their default hyper-parameters.
 
\subsection{Baseline Methods}
We compare \gear with two baseline methods:

\begin{itemize}[leftmargin=*,noitemsep,topsep=0pt]
\item \emph{Turn-control:} Gao and Peng~\cite{gao2026more} recently propose a turn-control approach to reduce the cost of SE agents, where a turn denotes a single tool invocation. The agent is constrained by an initial turn limit and granted a one-time extension if the limit is reached before completion. Following the original work, we set the initial limit to the 25th percentile of turn counts observed on SWE-bench Lite, and extend it to the 50th percentile if needed.

\item \emph{Naive-\gear:} RAG has been widely adopted to improve SE agents by retrieving relevant context to augment generation~\cite{abs251004905}. To isolate the effect of experience abstraction in \gear, we implement a variant, termed Naive-\gear, which retrieves raw execution trajectories and generated patches as context, instead of the structured experience objects used in \gear. All other components and settings are kept identical to those of \gear.

\end{itemize}

\subsection{Evaluation Metrics}
We evaluate \gear in terms of both task performance and cost efficiency.

\noindent \textbf{Performance Metric.} We use the resolution rate (denoted as \emph{\% Resolved}) as the metric, defined as the percentage of tasks for which an agent produces a patch that passes all tests~\cite{openai2025swebenchverified}.

\noindent \textbf{Efficiency Metrics.}

We comprehensively measure cost efficiency from multiple perspectives:
\begin{itemize}[leftmargin=*,noitemsep,topsep=0pt]
\item \emph{\# API Calls:} The total number of LLM API invocations across the benchmark.

\item \emph{\# Input/Output Tokens:} The total number of input and output tokens consumed by LLMs across the benchmark.

\item \emph{Total Cost:}  The total monetary cost incurred across the benchmark, measured in U.S. dollars.

\end{itemize}

\subsection{Implementation Details}

To tune the hyper-parameters of \gear, including $\tau_{\text{sim}}$, $\tau^{\text{gen}}$, $\tau_{\text{upper}}^{\text{sel}}$, and $\tau_{\text{lower}}^{\text{sel}}$, we randomly select 100 issues from the SWE-bench dataset~\cite{jimenez2024swebench} that are not included in SWE-bench Verified or SWE-bench Lite to serve as validation data. Detailed hyper-parameter tuning procedures are provided in Appendix~\ref{hptuning}.
\definecolor{upblue}{RGB}{0, 112, 192}
\definecolor{downorange}{RGB}{237, 125, 49}

\newcommand{\perf}[5]{#1 $\rightarrow$ #2 {\scriptsize\textbf{\color{#5}(#3#4)}}}

\begin{table}[t]
\centering
\small
\begin{tabular}{l r r}
\toprule
\textbf{Agent} & \textbf{GPT-5-mini} & \textbf{DeepSeek-V3.2} \\
\midrule
Agentless      & 14.0\% & 11.8\% \\
Mini-SWE-Agent & 10.6\% &  8.6\% \\
Trae Agent     & 12.4\% & 10.2\% \\
\midrule
\midrule
\textbf{Average} & \multicolumn{2}{c}{\textbf{11.3\%}}\\
\bottomrule
\end{tabular}
\caption{(RQ1) Early termination rates of different agents with \gear.}
\label{tab:early_termination}
\end{table}

\begin{table*}[h]
\small
  \centering
  
  \begin{tabular}{lrrrrr}
    \toprule
    \textbf{Method} & \textbf{\% Resolved} & \textbf{\# API Calls} & \textbf{\# Input Tokens} & \textbf{\# Output Tokens} & \textbf{Total Cost} \\
    \midrule
    Turn-control& -10.7\% & -33.9\% & -45.3\% & -31.8\% & -41.4\% \\
    Naive-\gear & -3.1\% & -16.6\% & -22.7\% & -20.8\% & -26.0\% \\
    \gear & +2.7\% & -20.8\% &  -29.9\%  & -25.1\%  & -31.8\% \\
    \bottomrule
  \end{tabular}
  \caption{(RQ2) Average impact of \gear and the two baseline methods on task performance and cost across the evaluated SE agents.}
  \label{tab:baseline_results}
\end{table*}

\section{Results}

\subsection{RQ1: Effectiveness of \gear}
This RQ evaluates the effectiveness of \gear in terms of its impact on both task performance and cost. Table~\ref{tab:mainResults} compares performance and cost with and without \gear across six agent configurations, instantiated by Agentless, Mini-SWE-Agent, and Trae Agent with two LLM backends (i.e., GPT-5-mini and DeepSeek-V3.2).  
Overall, \gear substantially reduces cost while preserving and, in some cases, improving task performance.

\noindent \textbf{Cost analysis.} \gear reduces total cost by 19.3\%–55.1\% across the six configurations, with an average reduction of 31.8\%. This cost saving is consistently observed across different agent paradigms and LLM backends. For Mini-SWE-Agent, the cost reduction remains significant but is smaller than that of the other two agents. This is because \gear affects only the cost of generating a single patch for Mini-SWE-Agent, whereas for the other agents it can reduces cost by avoiding repeated patch generation–selection cycles, resulting in larger overall savings.

Cost reductions are also consistent across multiple dimensions. Specifically, \gear reduces the number of API calls by 7.9\%-29.9\%, input tokens by 13.6\%-51.8\%, and output tokens by 3.7\%- 51.0\%, with average reductions of 20.8\%, 29.9\%, and 25.1\%, respectively. 

Examining cost from another perspective, we consider the early termination rate of issues. Table~\ref{tab:early_termination} presents the results: with \gear, the agents achieve early termination for an average of 11.3\% of issues, ranging from 8.6\% to 14.0\% across different agents.

Furthermore, \gear effectively reduces cost in both patch selection and patch generation. Detailed results and analysis are provided in Appendix~\ref{finegrained}. We further provide three detailed case studies in Appendix~\ref{appendix:casestudy} that illustrate how \gear guides early termination on concrete issues.

The effectiveness of \gear relies on the raw confidence scores emitted by the LLM. We find that these scores are well-calibrated: patches with confidence $>90$ achieve pass rates of 63.6\%–92.6\% across agent-model combinations, whereas patches with confidence $<40$ have pass rates of only 8.7\%–13.8\% (see Appendix~\ref{appendix:calibration} for details). This calibration provides the empirical basis for using confidence as a reliable termination signal.

Additionally, to assess the cross-repository transferability of \gear, we evaluate Trae Agent on 50 issues whose repositories are entirely absent from the experience base. With zero repository overlap, \gear still reduces cost from \$12.7 to \$9.6 while maintaining the resolution rate at 14.0\%, indicating that the retrieved experience captures generalizable debugging patterns rather than repository-specific cues. Detailed results are provided in Appendix~\ref{appendix:crossrepo}.

\noindent \textbf{Performance analysis.} Across the six agent configurations, \gear maintains task performance with negligible loss. Only a minor decrease of 0.2\% is observed for Trae Agent with DeepSeek-V3.2; however, McNemar's test~\cite{mcnemar1947note} indicates that this difference is not statistically significant ($p$ = 0.460). For Trae Agent with GPT-5-mini and Mini-SWE-Agent with both GPT-5-mini and DeepSeek-V3.2, resolution rates remain nearly unchanged after applying \gear. Notably, for Agentless, \gear increases the resolution rate by 7.8\% and 7.2\% with GPT-5-mini and DeepSeek-V3.2, respectively. This improvement is likely because Agentless exhibits relatively lower resolution rates, allowing the experience-driven guidance in \gear to have a more substantial impact on performance.

\finding{\gear substantially reduces total cost by 19.3\%–55.1\% (average 31.8\%) across the evaluated SE agents, with negligible loss in resolution rate (at most 0.2\%). In particular, it reduces API calls by 7.9\%–29.9\% (average 20.8\%), input tokens by 13.6\%–51.8\% (average 29.9\%), and output tokens by 3.7\%–51.0\% (average 25.1\%). It also achieves early termination for 8.6\% to 14.0\% of issues (average 11.3\%).}

\subsection{RQ2: Comparison with Baselines}
This RQ compares \gear with two representative baseline methods: Turn-control and Naive-\gear. As described in Section~\ref{agentdes}, Agentless follows a predefined, fixed workflow for patch generation and does not support autonomous tool invocation; therefore, turn-control, which explicitly limits tool invocation turns, cannot be applied to Agentless.

To save space, Table~\ref{tab:baseline_results} reports the average effects of the three methods on task performance and cost across the evaluated SE agents, while detailed results for each agent are provided in Table~\ref{tab:baseline_results_detail}.

\begin{table*}[h]
  \centering
  \small
  \begin{tabular}{lrrrrr}
    \toprule
    \textbf{Method} & \textbf{\% Resolved} & \textbf{\# API Calls} & \textbf{\# Input Tokens} & \textbf{\# Output Tokens} & \textbf{Total Cost} \\
    \midrule
    \gear & 0.0\% & -29.9\% & -30.4\% & -28.0\% & -28.2\% \\
    w/o experience injection & -10.4\% & -41.9\% & -41.7\% & -40.7\% & -58.9\% \\
    w/o early termination & +0.4\% & +9.2\% & +7.3\% & +8.0\% & +3.1\% \\
    \bottomrule
  \end{tabular}
  \caption{(RQ3) Impact of \gear and its ablated variants on task performance and cost for Trae Agent with GPT-5-mini.}
  \label{tab:ablation}
\end{table*}

\noindent \textbf{Comparison with Turn-control.} While Turn-control achieves a larger reduction in cost, it does so at the expense of substantial task performance loss. On average, Turn-control reduces total cost by 41.4\%, compared to 31.8\% achieved by \gear; however, it also decreases resolution rate by 10.7\%, whereas \gear improves resolution rate by 2.7\% on average. This cost-performance pattern is consistently observed across all agent configurations evaluated, as shown in Table~\ref{tab:baseline_results_detail}.

These results indicate that \gear strikes a superior balance between cost and task performance. This advantage stems from \gear’s ability to dynamically decide whether to terminate iterations early based on the agent’s current execution status and previously collected experience, allowing it to avoid unnecessary iterations without prematurely stopping productive ones.

\noindent \textbf{Comparison with Naive-\gear.} Overall, \gear outperforms Naive-\gear in both cost reduction and task performance preservation. 

While Naive-\gear also leverages previous execution trajectories to enable early termination and reduce cost, these raw trajectories are larger and less structured than the experience objects used by \gear, resulting in higher overhead. Thus, \gear achieves a greater cost reduction. In addition, the unstructured nature of the raw trajectories provides less effective guidance for patch generation and selection, whereas \gear’s structured experience objects better support decision-making, leading to improved task performance.

\finding{Overall, \gear achieves a better balance between cost reduction and task performance: while Turn-control reduces cost more aggressively, it incurs a substantial performance loss (10.7\% on average), and \gear outperforms Naive-\gear in both cost savings and performance preservation.}

\subsection{RQ3: Ablation Study}
This RQ investigates the individual contributions of experience injection and early termination in \gear. To this end, we conduct an ablation study with two variants: (1) disabling experience injection while retaining early termination, and (2) disabling early termination in patch generation and selection while retaining experience injection. Due to budget constraints, this experiment is performed only on Trae Agent, the most advanced agent among those evaluated. We use GPT-5-mini as the backend because it incurs lower inference cost than DeepSeek-V3.2.
Table~\ref{tab:ablation} shows the results.

\noindent \textbf{Without experience injection.} When experience injection is removed, the resolution rate decreases by 10.4 percentage points compared to the original \gear, although API calls, input tokens, output tokens, and total cost decrease by an additional 12.0\%, 11.3\%, 12.7\%, and 30.7\%, respectively.
The performance degradation is expected. Without experience, early termination relies solely on local or heuristic signals and may prematurely terminate promising patch generation or selection processes. Experience injection provides historical guidance that helps distinguish unproductive iterations from those likely to yield valid patches. Removing this guidance increases the risk of terminating effective search paths, leading to lower task performance, even though cost is further reduced due to the absence of experience-related token overhead.

\noindent \textbf{Without early termination.} In contrast, when early termination is disabled, the resolution rate increases 0.4 percentage points, while API calls, input tokens,  output tokens, and total cost increase by 9.2\%, 7.3\%, 8.0\% and 3.1\%, respectively. This behavior is also expected: disabling early termination allows the agent to explore more iterations during patch generation and selection, increasing the likelihood of finding valid patches. However, this comes at the expense of higher cost, as experience-related token overhead is incurred without the compensating benefit of early stopping. 

\finding{Removing experience injection leads to a substantial drop in resolution rate, despite further reducing cost. In contrast, disabling early termination slightly improves resolution rate, but leads to increased cost. 
}

\section{Conclusion}
We present \gear, an experience-driven early termination approach that reduces the cost of SE agents while preserving task performance. By leveraging structured experience from prior executions and applying experience-guided early termination during patch generation and selection, \gear mitigates cost inefficiencies caused by redundant iterations in current SE agents. Our evaluation across multiple agent paradigms and LLM backends demonstrates consistent cost reductions (31.8\% on average) with negligible loss on resolution rate (at most 0.2\%). These gains stem from identifying opportunities for early termination, reducing unnecessary API calls, and limiting token usage.
Importantly, \gear is a highly general optimization approach that can be integrated seamlessly into diverse agents without requiring fundamental redesigns.

\section*{Limitations}
While \gear demonstrates significant cost reductions and maintains task performance, this paper has several limitations.  

\noindent \emph{(1) Evaluation scope.} Consistent with prior SE agent studies~\cite{gao2025trae,gao2026more}, our evaluation is limited to SWE-bench Verified, a human-validated benchmark released by OpenAI for reliably assessing AI models’ ability to resolve SE issues. In future work, we plan to extend the evaluation to additional benchmarks as more reliable datasets become available. If feasible, we also aim to evaluate \gear in industrial settings. 

\noindent \emph{(2) Dependence on historical data.} As an experience-driven approach, \gear relies on historical data to construct its experience base. For entirely novel issues or domains with sparse historical data, i.e., the ``cold start'' problem, \gear may fail to retrieve relevant experience. This limitation is common to RAG-based approaches and motivates further exploration of methods that generalize better in low-data scenarios.

\noindent \emph{(3) Domain scope.} We focus on SE agents because they provide a realistic and cost-sensitive multi-step reasoning setting for evaluation. Nonetheless, the design philosophy of \gear is domain-agnostic: it operates at the level of context generation and budget-aware execution rather than task-specific logic. Extending \gear to non-programming domains (e.g., general multi-step reasoning agents) is an important direction for future work.

\section*{Acknowledgments}
Yaoqi Guo and Yang Liu are supported by Ministry of Defence (MINDEF), Singapore, through the Defence Science and Technology Agency (DSTA) under the Project Agreement number 9025202776. Any opinions, findings, conclusions, or recommendations expressed in these materials are those of the author(s) and do not reflect the views of the MINDEF/DSTA. Jie M. Zhang is supported by the ITEA grants GreenCode (project number 23016) and GENIUS (project number 23026). 

\bibliography{custom}
\appendix

\renewcommand{\topfraction}{0.95}
\renewcommand{\bottomfraction}{0.85}
\renewcommand{\textfraction}{0.05}
\renewcommand{\floatpagefraction}{0.85}
\setcounter{topnumber}{3}
\setcounter{bottomnumber}{2}
\setcounter{totalnumber}{5}

\section{Experience Object Example}
\label{appendix:experience_example}

\begin{figure*}[!t]
    \centering
    \includegraphics[width=0.65\linewidth]{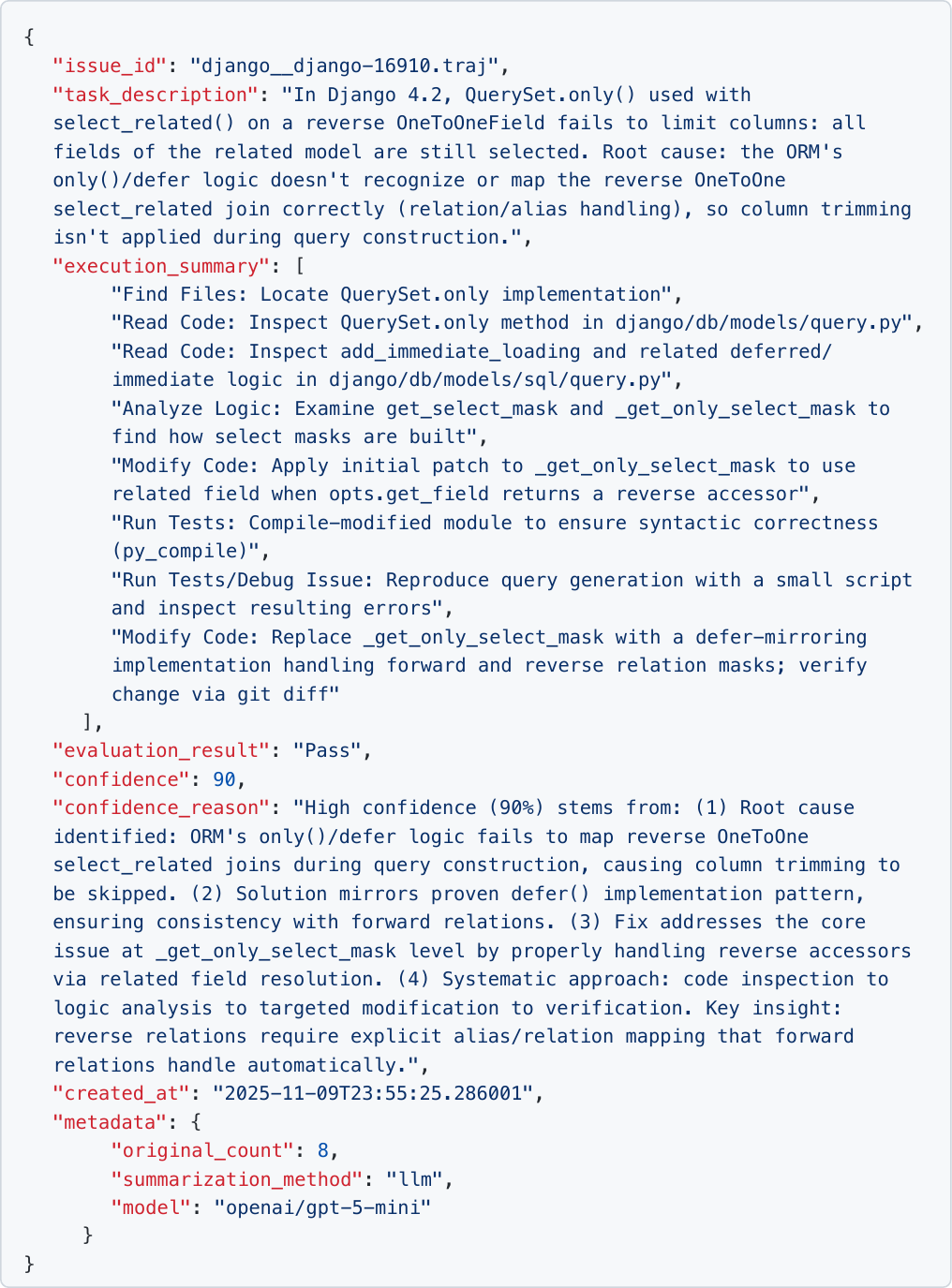}
    \caption{Experience object of the django\_\_django-16910 issue.}
    \label{appfig:experience_example}
\end{figure*}

Figure \ref{appfig:experience_example} shows an example experience object for the django\_\_django-16910 issue, constructed from the execution of Mini-SWE-Agent with GPT-5-mini.

\section{Hyper-Parameter Tuning of \gear}\label{hptuning}
\gear involves three hyper-parameters, including $\tau_{\text{sim}}$, $\tau^{\text{gen}}$, $\tau_{\text{lower}}^{\text{sel}}$, and $\tau_{\text{upper}}^{\text{sel}}$. To tune them, we randomly select 100 issues from the SWE-bench dataset~\cite{jimenez2024swebench} that are not included in SWE-bench Verified or SWE-bench Lite to serve as validation data.

\noindent \textbf{Tuning of $\tau_{\text{sim}}$.} The similarity threshold $\tau_{\text{sim}}$ controls the quality of experience retrieval in \gear. A threshold that is too low may retrieve irrelevant experience objects, while a threshold that is too high may retrieve too few, rendering \gear ineffective due to insufficient guidance. To identify an appropriate value, we conduct tuning experiments on the validation data using Mini-SWE-Agent with GPT-5-mini, chosen for its fast execution among the evaluated agents. During this tuning, we apply experience injection without early termination. The threshold is varied from 0.1 to 0.3 in steps of 0.05. 

\begin{table}[!htbp]
\small
\centering
\begin{tabular}{rrr}
\toprule
$\tau_{\text{sim}}$ & \% Resolved & Cost \\ \midrule
0.10        & 18\%        & \$2.98 \\
0.15        & 24\%        & \$3.16 \\
0.20        & 24\%        & \$3.18 \\
0.25        & 20\%        & \$3.49 \\
0.30        & 18\%        & \$4.06 \\ \bottomrule
\end{tabular}
\caption{Estimated resolution rate and total cost for 100 issues with stochastic perturbations.}
\label{tab:choosing_similarity}
\end{table}

Table~\ref{tab:choosing_similarity} reports the resolution rate and total cost for each threshold. As shown, $\tau_{\text{sim}} = 0.15$ achieves both the highest resolution rate and the lowest cost, and is thus adopted for all subsequent evaluations.

\noindent \textbf{Tuning of $\tau^{\text{gen}}$.} The threshold $\tau^{\text{gen}}$ controls the early termination of patch generation. Setting $\tau^{\text{gen}}$ too high limits the effectiveness of early stopping, whereas setting it too low may terminate generation prematurely and degrade task performance. To identify an appropriate value, we conduct tuning experiments on the validation data using Mini-SWE-Agent with GPT-5-mini, chosen for its fast execution among the evaluated agents. The threshold is varied from 70 to 90 in increments of 5, and we also evaluate the case without applying early termination.

\begin{table}[!htbp]
\small
\centering
\begin{tabular}{rrr}
\toprule
$\tau^{\text{gen}}$ & \% Resolved & Cost \\ \midrule
70                  & 8\%         & \$2.54                  \\
75                  & 8\%         & \$2.61                  \\
80                  & 12\%        & \$2.84                  \\
85                  & 20\%        & \$2.95                  \\
90                  & 24\%        & \$3.10                  \\
w/o                 & 24\%        & \$3.16                  \\ \bottomrule
\end{tabular}
\caption{Estimated resolution rate and total cost for 100 issues across different $\tau^{\text{gen}}$ settings with stochastic perturbations.}
\label{tab:choosing_confidence_1}
\end{table}

The results, summarized in Table~\ref{tab:choosing_confidence_1}, indicate that a threshold of 90 achieves cost reduction while maintaining the resolution rate comparable to that of no early termination. Accordingly, we adopt $\tau^{\text{gen}} = 90$ for all subsequent experiments.

\noindent \textbf{Tuning of $\tau_{\text{lower}}^{\text{sel}}$ and $\tau_{\text{upper}}^{\text{sel}}$.} The two thresholds control early termination during patch selection. Specifically, $\tau_{\text{upper}}^{\text{sel}}$ defines the confidence boundary above which a patch is assumed correct and no further generation is needed, while $\tau_{\text{lower}}^{\text{sel}}$ defines the boundary below which the issue is considered too difficult to resolve completely.

Unlike the previous thresholds, tuning of these two is conducted on the validation set using Trae Agent with GPT-5-mini, since Mini-SWE-Agent does not include a patch selection phase. For $\tau_{\text{lower}}^{\text{sel}}$, we vary its value from 0 to 50 in increments of 10, leaving $\tau_{\text{upper}}^{\text{sel}}$ unset during this tuning. For $\tau_{\text{upper}}^{\text{sel}}$, we vary it from 80 to 100 in increments of 5, leaving $\tau_{\text{lower}}^{\text{sel}}$ unset.

\begin{table}[!htbp]
\small
  \centering
  \begin{tabular}{rrr}
    \toprule
    Confidence & \% Resolved & Cost \\
    \midrule
    0  & 28\% & \$34.8 \\
    10 & 28\% & \$34.6 \\
    20 & 28\% & \$33.9 \\
    30 & 28\% & \$31.3 \\
    40 & 28\% & \$28.2 \\
    50 & 24\% & \$26.1 \\
    \bottomrule
  \end{tabular}
  \caption{Resolution rate and total cost across different $\tau_{\text{lower}}^{\text{sel}}$ settings.}
  \label{tab:confidence_lower}
\end{table}

\begin{table}[!htbp]
\small
  \centering
  \begin{tabular}{rrr}
    \toprule
    Confidence & \% Resolved & Cost \\
    \midrule
    100 & 28\% & \$34.8 \\
    95  & 28\% & \$33.1 \\
    90  & 28\% & \$31.2 \\
    85  & 25\% & \$30.2 \\
    80  & 21\% & \$28.6 \\
    \bottomrule
  \end{tabular}
  \caption{Resolution rate and total cost across different $\tau_{\text{upper}}^{\text{sel}}$ settings.}
  \label{tab:confidence_higher}
\end{table}

Tables~\ref{tab:confidence_lower} and~\ref{tab:confidence_higher} summarize the results. Based on these experiments, we select $\tau_{\text{lower}}^{\text{sel}} = 40$, which achieves the highest resolution rate at the lowest cost, and $\tau_{\text{upper}}^{\text{sel}} = 90$, which similarly balances maximal resolution with minimal cost.

In patch selection, the higher threshold determines the boundary where the patch is assumed to be correct and needs no regeneration, and the lower threshold determines the boundary where the issue is too hard to complete. To determine the thresholds, we use the same subset of SWE-bench and do experiment on Trae-agent with GPT-5-mini as backend. According to Table \ref{tab:confidence_lower} and \ref{tab:confidence_higher}, we choose 40 as the lower threshold and 90 as the higher threshold, for they are the thresholds that minimize cost without compromising performance.

\section{Cost Reduction in Patch Generation and Selection}\label{finegrained}
To conduct a fine-grained cost analysis of \gear, we examine its impact on patch generation and patch selection separately. For patch generation, we measure the number of API calls per patch before and after applying \gear; for patch selection, we compute the average number of generated patches before and after applying \gear. All analyses are performed separately for resolved and unresolved issues.

Table~\ref{tab:api_iters_passfail} reports the number of API calls per patch for resolved and unresolved issues, before and after applying \gear. Since Agentless follows a fixed workflow to generates individual patches, we exclude it from this analysis. With the exception of Mini-SWE-Agent using GPT-5-mini, \gear reduces the number of API calls per patch across all other configurations. Notably, the reduction is more pronounced for unresolved issues than for resolved ones. For example, for Mini-SWE-Agent with DeepSeek V3.2, applying \gear reduces the API calls per patch from 61.7 to 58.9 for resolved issues (a decrease of 2.8), and from 79.5 to 68.7 for unresolved issues (a decrease of 10.8). These results indicate that \gear effectively avoids prolonged iterations on difficult or unsolvable issues, which would otherwise continue to incur API calls with limited benefit to successful issue resolution, thereby significantly reducing unnecessary cost.

\begin{table}[!htbp]
\centering
\small
\begin{tabular}{lcc}
\toprule
\textbf{Agent} & \makecell[r]{\textbf{\# API Calls}\\\textbf{(Resolved)}} & \makecell[r]{\textbf{\# API Calls}\\\textbf{(Unresolved)}} \\
\midrule

\multicolumn{3}{c}{\textbf{GPT-5-mini}} \\
\midrule
Mini-SWE-Agent & 13.0 & 16.4 \\
\quad \textit{with \gear}        & 13.0 & 13.8 \\
\midrule

\multicolumn{3}{c}{\textbf{DeepSeek V3.2}} \\
\midrule
Mini-SWE-Agent & 61.7 & 79.5 \\
\quad \textit{with \gear}        & 58.9 & 68.7 \\
\midrule

\multicolumn{3}{c}{\textbf{GPT-5-mini}} \\
\midrule
Trae Agent      & 36.2 & 44.7 \\
\quad \textit{with \gear}  & 36.1  & 39.9  \\
\midrule
\multicolumn{3}{c}{\textbf{DeepSeek V3.2}} \\
\midrule
Trae Agent      & 98.8 & 105.1 \\
\quad \textit{with \gear}       & 97.4 & 101.5 \\
\bottomrule
\end{tabular}
\caption{Number of API calls per patch for resolved and unresolved issues, before and after applying \gear.}
\label{tab:api_iters_passfail}
\end{table}

Table~\ref{tab:patch_count_passfail} reports the average number of patches generated for resolved and unresolved issues, before and after applying \gear, for agents that perform patch selection (i.e., Agentless and Trae Agent). The original Agentless and Trae Agent use a fixed number of generated patches. In contrast, \gear adaptively reduces the number of patches for both resolved and unresolved issues across all configurations. For example, with GPT-5-mini, Agentless originally generates four patches, whereas \gear reduces this to 2.0 patches for resolved issues and 2.7 patches for unresolved issues. These results demonstrate that \gear effectively reduces patch-generation iterations.

\begin{table}[!htbp]
\centering
\small
\begin{tabular}{lcc}
\toprule
\textbf{Agent} & \makecell[r]{\textbf{\# Patches}\\\textbf{(Resolved)}} & \makecell[r]{\textbf{\# Patches}\\\textbf{(Unresolved)}} \\
\midrule

\multicolumn{3}{c}{\textbf{GPT-5-mini}} \\
\midrule
Agentless & 4.0 & 4.0 \\
\quad \textit{with \gear}       & 2.0 & 2.7 \\
\midrule

\multicolumn{3}{c}{\textbf{DeepSeek V3.2}} \\
\midrule
Agentless & 4.0 & 4.0 \\
\quad \textit{with \gear} & 1.9 & 2.7 \\
\midrule

\multicolumn{3}{c}{\textbf{GPT-5-mini}} \\
\midrule
Trae Agent  & 3.0 & 3.0 \\
\quad \textit{with \gear}  & 2.1 & 2.3 \\
\midrule
\multicolumn{3}{c}{\textbf{DeepSeek V3.2}} \\
\midrule
Trae Agent     & 3.0 & 3.0 \\
\quad \textit{with \gear}  & 2.1 & 2.5 \\
\bottomrule
\end{tabular}
\caption{Average number of generated patches for resolved and unresolved issues, before and after applying \gear.}
\label{tab:patch_count_passfail}
\end{table}

\begin{table*}[!t]
\small
  \centering
  
  \begin{tabular}{lrrrrr}
    \toprule
    \textbf{Method} & \textbf{\% Resolved} & \textbf{\# API Calls} & \textbf{\# Input Tokens} & \textbf{\# Output Tokens} & \textbf{Total Cost} \\
    
    \midrule
    \multicolumn{6}{c}{\textbf{Agentless + GPT-5-mini}} \\
    \midrule
    Naive-\gear & +1.0\% & -24.0\% & -48.7\% & -49.9\% & -53.0\% \\
    \gear & +7.8\% & -26.4\% & -51.8\% & -51.0\% & -55.1\% \\
    
    \midrule
    \multicolumn{6}{c}{\textbf{Agentless + DeepSeek-V3.2}} \\
    \midrule
    Naive-\gear & +1.2\% & -21.8\% & -26.3\% & -28.7\% & -25.8\% \\
    \gear & +7.2\% & -25.5\% & -31.9\% & -35.0\% & -31.8\% \\
    
    \midrule
    \multicolumn{6}{c}{\textbf{Mini-SWE-Agent + GPT-5-mini}} \\
    \midrule
    Turn-control & -13.0\% & -33.5\% & -60.1\% & -38.1\% & -50.5\% \\
    Naive-\gear & -0.6\% & +2.7\% & +16.0\% & +17.0\% & -1.1\% \\
    \gear & +1.0\% & -7.9\% & -13.7\% & -3.7\% & -19.4\% \\
    
    \midrule
    \multicolumn{6}{c}{\textbf{Mini-SWE-Agent + DeepSeek-V3.2}} \\
    \midrule
    Turn-control & -11.0\% & -28.4\% & -46.8\% & -17.1\% & -41.6\% \\
    Naive-\gear & -4.2\% & -4.2\% & -8.5\% & -4.0\% & -8.2\% \\
    \gear & +0.6\% & -8.4\% & -13.6\% & -4.4\% & -19.3\% \\
    
    \midrule
    \multicolumn{6}{c}{\textbf{Trae Agent + GPT-5-mini}} \\
    \midrule
    Turn-control & -10.0\% & -34.6\% & -34.9\% & -32.5\% & -34.6\% \\
    Naive-\gear & -9.0\% & -34.1\% & -34.7\% & -32.1\% & -31.7\% \\
    \gear & 0.0\% & -29.9\% & -30.4\% & -28.0\% & -28.2\% \\
    
    \midrule
    \multicolumn{6}{c}{\textbf{Trae Agent + DeepSeek-V3.2}} \\
    \midrule
    Turn-control & -8.8\% & -39.0\% & -39.5\% & -39.5\% & -38.7\% \\
    Naive-\gear & -7.0\% & -18.2\% & -34.0\% & -27.0\% & -36.4\% \\
    \gear & -0.2\% & -26.5\% & -37.7\% & -28.2\% & -36.7\% \\
    \midrule
    \multicolumn{6}{c}{\textbf{Average}} \\
    \midrule
    Turn-control& -10.7\% & -33.9\% & -45.3\% & -31.8\% & -41.4\% \\
    Naive-\gear & -3.1\% & -16.6\% & -22.7\% & -20.8\% & -26.0\% \\
    \gear & +2.7\% & -20.8\% &  -29.9\%  & -25.1\%  & -31.8\% \\
    \bottomrule
  \end{tabular}
  \caption{(RQ2) Impact of \gear and baseline methods on SE agents' task performance and cost.}
  \label{tab:baseline_results_detail}
\end{table*}

\section{Detailed Comparison with Baselines}
Table~\ref{tab:baseline_results_detail} presents the effects of \gear and the two baseline methods on task performance and cost across different SE agents.

\section{Case Studies}\label{appendix:casestudy}

To illustrate how \gear influences agent behavior in practice, we present three representative cases. Each highlights a distinct mechanism through which retrieved experience contributes: guided bug localization (Case 1), avoidance of a superficially plausible but incorrect fix (Case 2), and confidence-based early termination of an already-correct patch (Case 3).

\paragraph{Case 1: Mini-SWE-Agent on \texttt{django-11211}.} The issue arises because \texttt{prefetch\_related} fails when a \texttt{GenericForeignKey} points to a \texttt{UUIDField} primary key: the foreign key stores a string while the primary key is a UUID, so matching fails. Without \gear, the agent spends multiple steps surveying unrelated parts of the Django ORM before locating the fault. With a retrieved experience entry indicating \texttt{GenericForeignKey.get\_prefetch\_queryset} as the likely failure site, the agent goes directly to that function and produces a patch that normalizes the ID types before matching. In this case, the benefit of experience injection is reduced exploration overhead rather than a change in the final fix.

\paragraph{Case 2: Trae Agent on \texttt{astropy-12907}.} The issue concerns an incorrect separability matrix for nested \texttt{CompoundModels}, which causes spurious input/output coupling between otherwise independent sub-models. A surface-level remedy is to adjust the matrix indices directly, which masks the symptom but leaves the underlying recursion bug in place. With a retrieved experience describing the recursive tree structure of nested \texttt{CompoundModels}, Trae+\gear instead identifies premature termination in the recursion and produces a patch that unpacks sub-models so that the diagonal-block structure of the separability matrix is recovered. Here, the retrieved experience steers the agent away from a symptomatic fix toward the root cause.

\paragraph{Case 3: Agentless on \texttt{xarray-7233}.} In this issue, \texttt{coarsen.construct} demotes non-dimensional coordinates to regular data variables. Agentless's first generates patch restored the coordinate status by reusing the original dataset's coordinate keys. Conditioned on the retrieved experience, \gear's confidence estimator assigns this patch a score high enough to trigger early termination, avoiding the generation of additional candidate patches. Compared to Cases 1 and 2, the saving here comes from avoided downstream work rather than improved localization.

Taken together, these cases show that \gear's benefits surface through distinct mechanisms depending on the agent and the issue: faster localization, redirection away from symptomatic fixes, and early termination once a correct patch is already in hand.

\section{Cross-Repository Transferability}\label{appendix:crossrepo}

An important question for experience-driven approaches is whether the observed cost reduction reflects transferable debugging logic or project-specific familiarity. Although overlapping tasks between SWE-bench Lite (used for experience generation) and SWE-bench Verified (used for evaluation) are removed, both benchmarks may still share repositories. To examine this question more rigorously, we randomly select 50 issues from SWE-bench whose repositories are entirely absent from all existing experience bases. We use Trae Agent for this evaluation and adopt GPT-5-mini as the backbone model.

\begin{table}[!htbp]
\small
  \centering
  \begin{tabular}{lrr}
    \toprule
    \textbf{Agent} & \textbf{\% Resolved} & \textbf{Total Cost} \\
    \midrule
    Trae  Agent      & 14\% & \$12.7 \\
    \quad \textit{with \gear} & 14\% & \$9.6 \\
    \bottomrule
  \end{tabular}
  \caption{Cross-repository transferability of \gear. Evaluation is performed on 50 issues whose repositories are absent from all experience bases.}
  \label{tab:crossrepo}
\end{table}

As shown in Table~\ref{tab:crossrepo}, even with zero repository overlap, \gear still reduces the average cost from \$12.7 to \$9.6 (a 24.4\% reduction) while keeping the resolution rate unchanged at 14.0\%. This indicates that the retrieved experience captures generalizable debugging patterns rather than relying on memorized project-specific cues, confirming that \gear transfers robustly from known to unknown repositories.

\section{Scalability with Experience Base Size}\label{appendix:scale}

We further investigate how the effectiveness of \gear scales with the size of the experience base. Using Mini-SWE-Agent with GPT-5-mini, we conduct experiments on 50 issues out of the 100 issues used for hyper-parameter tuning while varying the number of entries in the experience base from 50 to the full set. Table~\ref{tab:scale} reports the resulting changes in resolution rate, cost reduction, and the number of experience injection times.

\begin{table}[!htbp]
\small
  \centering
  \begin{tabular}{lrrr}
    \toprule
    \textbf{\# Entries} & \textbf{\% Resolved} & \textbf{Cost Reduction} & \textbf{Injections} \\
    \midrule
    50   & +0\% & -8.0\%  & 8/50  \\
    100  & +0\% & -13.1\% & 11/50 \\
    150  & +2\% & -16.7\% & 15/50 \\
    all  & +2\% & -18.1\% & 24/50 \\
    \bottomrule
  \end{tabular}
  \caption{Effect of experience base size on \gear's solve rate, cost reduction, and experience injection frequency.}
  \label{tab:scale}
\end{table}

The results show that both cost reduction and the number of successful experience injections grow monotonically with the size of the experience base, while the resolution rate remains stable or slightly improves. This aligns with our expectation that \gear's benefit scales with the amount of accumulated historical data, supporting its long-term applicability as more issue-resolution trajectories are collected.

\begin{table*}[!t]
\small
\centering
\begin{tabular}{llrrr}
\toprule
\textbf{Agent} & \textbf{Difficulty} & \textbf{\% Resolved} & \textbf{Early Termination} & \textbf{Cost Reduction} \\
\midrule
\multicolumn{5}{c}{\textbf{GPT-5-mini}} \\
\midrule
\multirow{2}{*}{Agentless} & Simple  & 61.0\% & 16.3\% & -66.1\% \\
                           & Complex & 28.3\% & 12.8\% & -40.8\% \\
\multirow{2}{*}{Mini-SWE}  & Simple  & 72.2\% & 11.7\% & -20.6\% \\
                           & Complex & 46.7\% & 9.5\%  & -17.0\% \\
\multirow{2}{*}{Trae}      & Simple  & 82.3\% & 13.5\% & -33.7\% \\
                           & Complex & 56.0\% & 9.9\%  & -18.9\% \\
\midrule
\multicolumn{5}{c}{\textbf{DeepSeek-V3.2}} \\
\midrule
\multirow{2}{*}{Agentless} & Simple  & 63.4\% & 14.4\% & -48.9\% \\
                           & Complex & 41.2\% & 9.3\%  & -21.6\% \\
\multirow{2}{*}{Mini-SWE}  & Simple  & 81.4\% & 9.6\%  & -21.4\% \\
                           & Complex & 57.9\% & 7.9\%  & -18.0\% \\
\multirow{2}{*}{Trae}      & Simple  & 89.2\% & 11.4\% & -39.5\% \\
                           & Complex & 57.8\% & 9.1\%  & -32.5\% \\
\bottomrule
\end{tabular}
\caption{\gear's performance broken down by issue difficulty.}
\label{tab:difficulty}
\end{table*}

\section{Fine-Grained Analysis by Issue Difficulty}\label{appendix:difficulty}

To further understand how \gear behaves on issues of different difficulty, we categorize the 500 SWE-bench Verified issues using the official difficulty labels into 194 simple issues (requiring $<$15 minutes to fix) and 306 complex issues (requiring $\geq$15 minutes). Table~\ref{tab:difficulty} reports the per-category resolution rate, early termination rate, and cost reduction for each agent-model combination.

Overall, simple tasks achieve substantially higher resolution rates, early termination rates, and cost savings than complex tasks. This supports our design philosophy: \gear correctly lets agents spend fewer computational resources on easier tasks while still allocating sufficient iterations to harder ones.

\begin{table*}[!t]
\small
\centering
\begin{tabular}{llrrr}
\toprule
\textbf{Agent} & \textbf{Type} & \textbf{\% Resolved} & \textbf{Early Termination} & \textbf{Cost Reduction} \\
\midrule
\multicolumn{5}{c}{\textbf{GPT-5-mini}} \\
\midrule
\multirow{2}{*}{Agentless} & Bug Fix & 41.6\% & 14.1\% & -56.1\% \\
                           & Feature & 36.3\% & 12.8\% & -50.8\% \\
\multirow{2}{*}{Mini-SWE}  & Bug Fix & 57.3\% & 10.7\% & -19.5\% \\
                           & Feature & 51.1\% & 10.4\% & -19.0\% \\
\multirow{2}{*}{Trae}      & Bug Fix & 67.2\% & 12.5\% & -29.0\% \\
                           & Feature & 58.4\% & 10.9\% & -20.9\% \\
\midrule
\multicolumn{5}{c}{\textbf{DeepSeek-V3.2}} \\
\midrule
\multirow{2}{*}{Agentless} & Bug Fix & 50.4\% & 12.4\% & -34.2\% \\
                           & Feature & 45.1\% & 11.1\% & -29.8\% \\
\multirow{2}{*}{Mini-SWE}  & Bug Fix & 67.9\% & 8.6\%  & -19.4\% \\
                           & Feature & 60.0\% & 8.7\%  & -19.1\% \\
\multirow{2}{*}{Trae}      & Bug Fix & 70.7\% & 10.4\% & -37.0\% \\
                           & Feature & 64.6\% & 9.9\%  & -35.5\% \\
\bottomrule
\end{tabular}
\caption{\gear's performance broken down by issue type.}
\label{tab:issuetype}
\end{table*}

\section{Fine-Grained Analysis by Issue Type}\label{appendix:issuetype}

We further categorize the 500 SWE-bench Verified issues into 443 bug-fix issues and 57 feature-request issues. Table~\ref{tab:issuetype} reports the per-category resolution rate, early termination rate, and cost reduction for each agent-model combination.

Across all agents and backbones, bug-fix tasks exhibit higher resolution rates, higher early termination rates, and larger cost savings than feature-request tasks. This is consistent with the observation that bug-fix issues, being relatively simpler and more localized, are more amenable to confident early termination.

\begin{table*}[!t]
\small
\centering
\begin{tabular}{llr}
\toprule
\textbf{Agent} & \textbf{Model} & \textbf{Total Cost} \\
\midrule
\multirow{2}{*}{Agentless}      & GPT-5-mini    & \$4.3  \\
                                & DeepSeek-V3.2 & \$3.8  \\
\multirow{2}{*}{Mini-SWE-Agent} & GPT-5-mini    & \$7.1  \\
                                & DeepSeek-V3.2 & \$21.2 \\
\multirow{2}{*}{Trae Agent}     & GPT-5-mini    & \$22.5 \\
                                & DeepSeek-V3.2 & \$31.5 \\
\bottomrule
\end{tabular}
\caption{One-time, offline experience generation cost for \gear across agent-model combinations.}
\label{tab:expcost}
\end{table*}

\section{Experience Generation Cost}\label{appendix:expcost}

\gear's experience base is constructed once from historical executions and can be amortized over many downstream tasks, so its generation cost is not part of the per-task inference cost measured in RQ1 and RQ2. For transparency, we additionally report the one-time experience generation cost across all evaluated agent-model combinations in Table~\ref{tab:expcost}. The cost depends primarily on the verbosity of each agent's execution trajectories; autonomous agents such as Trae naturally incur higher generation cost than the lightweight Agentless.

\section{Top-$k$ Ablation on Retrieved Experiences}\label{appendix:topk}

We further study the effect of varying the number of retrieved experiences injected at each step. To this end, we conduct an ablation on the 50 issues used in Appendix~\ref{appendix:scale} using Mini-SWE-Agent with GPT-5-mini, varying the number of retrieved experiences $k$ from 1 to 3. Changes are reported relative to the baseline without \gear.

\begin{table}[!htbp]
\small
\centering
\begin{tabular}{rrr}
\toprule
\textbf{Top-$k$} & \textbf{\% Resolved} & \textbf{Total Cost} \\
\midrule
1 & +4\% & -18.1\% \\
2 & +0\% & -17.2\% \\
3 & -6\% & -15.6\% \\
\bottomrule
\end{tabular}
\caption{Effect of the number of retrieved experiences ($k$) on \gear's resolution rate and cost.}
\label{tab:topk}
\end{table}

As shown in Table~\ref{tab:topk}, injecting more than one experience entry gradually diminishes the effectiveness of \gear. This is likely due to the confusion and noise introduced by multiple competing historical trajectories, as well as the additional input-token cost. Combined with our observation that the most informative experiences are those derived from successful outcomes, we therefore adopt top-1 retrieval for all main experiments.

\section{Retrieval Method Comparison}\label{appendix:retrieval}

\gear uses TF-IDF as its default retrieval method due to its simplicity and low computational overhead. To examine the impact of this choice, we compare it against a modern embedding-based retriever using OpenAI's \texttt{text-embedding-3-small} on the same validation set. Table~\ref{tab:retrieval} reports relative changes over the no-retrieval baseline.

\begin{table}[!htbp]
\small
\centering
\begin{tabular}{lrr}
\toprule
\textbf{Retriever} & \textbf{\% Resolved} & \textbf{Total Cost} \\
\midrule
\texttt{text-embedding-3-small} & +0.6\% & -19.0\% \\
TF-IDF (default)                & +1.0\% & -19.4\% \\
\bottomrule
\end{tabular}
\caption{Comparison between the default TF-IDF retriever and an embedding-based retriever.}
\label{tab:retrieval}
\end{table}

\begin{table*}[!t]
\small
\centering
\begin{tabular}{llrrrr}
\toprule
\textbf{Agent} & \textbf{Model} & \textbf{$<$40} & \textbf{40--80} & \textbf{80--90} & \textbf{$>$90} \\
\midrule
\multirow{2}{*}{Agentless} & GPT-5-mini    & 8.7\%  & 24.8\% & 36.5\% & 63.6\% \\
                           & DeepSeek-V3.2 & 12.1\% & 22.9\% & 46.5\% & 78.2\% \\
\multirow{2}{*}{Trae}      & GPT-5-mini   & 13.8\% & 28.7\% & 47.0\% & 88.7\% \\
                           & DeepSeek-V3.2& 11.9\% & 33.2\% & 52.1\% & 92.6\% \\
\bottomrule
\end{tabular}
\caption{Pass rate of generated patches grouped by raw LLM confidence score.}
\label{tab:calibration}
\end{table*}

The two retrievers yield nearly identical results, with 212 out of 242 retrieved experience entries being identical across the two methods. Given this parity, we adopt TF-IDF as the lightweight default in \gear.

\section{LLM Confidence Calibration}\label{appendix:calibration}

The core mechanism of \gear hinges on the raw confidence score (0--100) emitted by the LLM when evaluating a generated patch. To verify that these scores provide a reliable signal for early termination, we bucket all generated patches by confidence and measure the actual pass rate in each bucket. Table~\ref{tab:calibration} reports the pass rate for four bucket ranges across representative agent-model combinations.

The results show a strong monotonic relationship between raw confidence and actual pass rate: patches with confidence scores above 90 achieve pass rates of 63.6\%--92.6\%, while those below 40 have pass rates of only 8.7\%--13.8\%. This empirically confirms that the raw confidence scores are reasonably well-calibrated and can serve as a reliable early-termination signal for \gear.

\section{License for Artifacts}
In this paper, we utilize the SWE-bench Verified benchmark~\cite{openai2025swebenchverified} (including the Lite subset for experience generation) and three open-source software agents: Agentless~\cite{xia2024agentless}, Mini-SWE-Agent~\cite{yang2024sweagent}, and Trae Agent~\cite{gao2025trae}. We confirm that all these artifacts are distributed under the \textbf{MIT License}. Additionally, we strictly adhere to the usage policies for the LLMs (GPT-5-mini and DeepSeek-V3.2) employed in our experiments.

\end{document}